\documentclass[twocolumn,prl,showpacs]{revtex4}

\usepackage{graphicx}

\begin{document}

\newcommand*{\cm}{cm$^{-1}$\/}
\newcommand{\comment}[1]{}
\newcommand*{\YBCO}{YBa$_2$Cu$_3$O$_{7-\delta}$\/}

\title{The Mysterious Pseudogap in High Temperature
Superconductors: An Infrared View.}

\author{T. Timusk}
\altaffiliation{The Canadian Institute of Advanced Research}
\email{timusk@mcmaster.ca}

\affiliation{Department of Physics and Astronomy, McMaster
University, Hamilton ON Canada, L8S 4M1}

\begin{abstract}

We review the contribution of infrared spectroscopy to the study of the
pseudogap in high temperature superconductors. The pseudogap appears as a
depression of the frequency dependent conductivity in the c-axis direction and
seems to be related to a real gap in the density of states. It can also be seen
in the Knight shift, photoemission and tunneling experiments. In underdoped
samples it appears near room temperature and does not close with temperature.
Another related phenomenon that has been studied by infrared is the depression
in the ab-plane scattering rate. Two separate effects can be discerned. At high
temperatures there is broad depression of scattering below 1000 \cm\ which may
be related to the gap in the density of states. At a lower temperature a
sharper structure is seen, which appears to be associated with scattering from
a mode at 300 \cm, and which governs the carrier life time at low temperatures.
This mode shows up in a number of other experiments, as a kink in ARPES
dispersion, and a resonance at 41 meV in magnetic neutron scattering. Since the
infrared technique can be used on a wide range of samples it has provided
evidence that the scattering mode is present in all high temperature cuprates
and that its frequency in optimally doped materials scales with the
superconducting transition temperature. The lanthanum and neodymium based
cuprates do not follow this scaling and appear to have depressed transition
temperatures.

\end{abstract}

\pacs{74.25.Kc, 74.25.Gz, 74.72.-h}

\maketitle

\section{Introduction}

Infrared spectroscopy has been used in superconductivity since the 1950's when
Tinkham  et al. first measured the size of the energy gap in lead with a prism
spectrometer using a mercury arc for a radiation source\cite{tinkham}. This was
well before the appearance of the BCS theory of superconductivity. The second
seminal experiment was the observation of phonon structure, again in the
reflectance spectra of strong coupling superconductor
lead\cite{joyce70,farnworth74,farnworth76,brandli72}, in 1972. By this time the
optical properties of the BCS superconductors were well
understood\cite{mattis58,allen71} and the phonon structure had been
predicted\cite{nam67}. The superconducting gap can only be observed in the
infrared in the dirty limit, $1/\tau \gg 2\Delta$ where $1/\tau$ is the elastic
scattering rate of the carriers and $\Delta$ the superconducting gap. In this
limit direct pairbreaking across the energy gap gives rise to a threshold in
absorption at $h\nu = 2\Delta$ where $\nu$ is the photon energy\cite{mattis58}
and a corresponding "knee" in reflectance spectra. In a clean limit BCS
superconductor there is also a knee in the reflectance spectrum but at a higher
frequency, $h\nu = 2\Delta+\hbar\Omega $ where $\Omega$ is the frequency of the
boson that is needed to conserve momentum in the absence of impurities. This
knee represents the onset of inelastic processes and the spectrum of bosons
responsible can be obtained by taking the second derivative of the reflectance
spectrum\cite{allen71,marsiglio98}. In a typical BCS superconductor there are
two peaks in this spectrum, one for transverse and one for longitudinal
phonons. The first derivative of the {\it difference} in reflectance between
the superconducting and normal states can also be used to extract the boson
spectral function\cite{farnworth74}.

%
\begin{figure}[t]
\vspace*{2.5cm}%
\centerline{\includegraphics[width=3.0in]{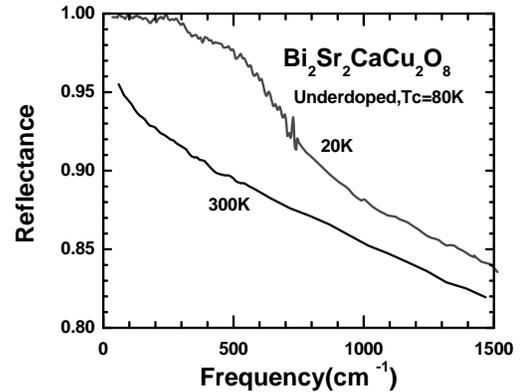}}%
\vspace*{-2.0cm}%
\caption{Infrared reflectance of a typical high temperature superconductor with
light polarized along the highly conducting ab-planes. At room temperature the
spectrum is Drude-like at low frequency but at high frequency the spectrum does
not flatten out, evidence of the presence of high frequency absorption
processes. At low temperature and low frequency the reflectance approaches
unity over a frequency scale of the order of 500 \cm suggestive of a gap-like
depression of scattering. In underdoped samples this gap appears well above the
superconducting transition temperature.}%
\label{isaksonfig1}
\end{figure}

The ab plane reflectance spectra of the high temperature superconductors also
exhibits a knee at $h\nu \approx 8k_BT_c$ at low temperature as shown in fig 1.
However, this is not a signature of the superconducting gap. First, the
scattering rate at low temperature is only $\approx k_BT_c$, which puts the
materials in the clean limit where the gap cannot be
seen,\cite{timusk88c,kamaras90} and second, at least in underdoped samples, the
knee already appears in the normal state\cite{reedyk88}. We now know that the
knee is a manifestation, not of the superconducting gap, but the result of a
complicated combination of the superconducting gap, the pseudogap and the
effect of the boson that seems to dominate transport scattering well above the
superconducting transition temperature
\cite{thomas88,basov96c,carbotte99,abanov01}.

The c-axis spectra are also dominated by a reflectance knee but it appears
strictly in the superconducting state. Unlike the ab plane knee it is not a
result of an onset of enhanced scattering but corresponds to a plasma edge,
very much like the plasma edge of silver in the ultraviolet\cite{bonn87}. This
has been identified as the Josephson plasma frequency\cite{tamasaku92}.

Kramers-Kronig analysis of the reflectance shows that the optical conductivity
is highly anisotropic, it is in the clean limit for transport along the
ab-plane and in the dirty limit for c-axis, interplane transport. This leads us
to an apparent contradiction: microwave ab-plane measurements\cite{bonn92} show
that at least YBCO crystals can be extremely pure with mean free paths
extending to microns rivaling the purest of conventional metals.  In the same
crystals, the c-axis transport is highly incoherent with mean free paths of
less than a unit cell. The poor interplane conductivity can not be caused by
defects but intrinsic incoherence. Incoherent transport occurs when the
probability of in-plane scattering is larger than the probability of interplane
charge transfer -- the electron has no time to set up a coherent interplane
wave function.

A complete review of the optical properties of high temperature superconductors
can be found in several recent papers\cite{basov01,timusk99}. The more limited
purpose of the present paper is to focus on the properties of the pseudogap as
seen by infrared spectroscopy and to update recent reviews. However, the
earliest evidence of a normal state gap in high temperature superconductors did
not come from the optical conductivity but from nuclear magnetic resonance
where  Warren et al.\cite{warren89} and Alloul et al.\cite{alloul89} found
evidence of a depressed density of states in the spin excitation spectrum. This
gap became known as the spin gap. The first {\it spectroscopic} evidence for a
pseudogap in high temperature superconductors came from infrared
spectroscopy\cite{homes93}. A pseudogap has been seen with many other
experimental probes. These include the Knight shift\cite{walstedt90}, dc
resistivity\cite{bucher93}, specific heat\cite{loram94}, angle resolved
photoemission (ARPES)\cite{loeser96,ding96,marshall96}, tunnelling
spectroscopy\cite{tao97,renner98,krasnov00},
Raman\cite{slakey90,hackl96,naeini99} and neutron
scattering\cite{rossat-mignod91,tranquada92}.

\section{The pseudogap in the c-axis conductivity}

Homes  et al.\cite{homes93}, in a study of the far infrared reflectivity of
c-axis oriented YBa$_2$Cu$_3$O$_{x}$ (YBCO), found, superimposed on an
otherwise flat background, a gap-like depression of conductivity, which
persisted up to room temperature. Fig 2, top panel, shows the raw data for the
c-axis optical conductivity obtained from a Kramers-Kronig transformation of
the reflectance of an $x=6.60$ sample. The spectrum is dominated by five strong
transverse optic phonons. The frequencies and eigenvectors of these phonons are
well known from neutron scattering\cite{timusk95}. The lower panel shows the
conductivity with the phonons subtracted. We note that at room temperature the
spectrum is flat within 5 \% . This high-temperature background conductivity
remains frequency and temperature independent at all doping levels up to
optimal doping at $x=6.95$\cite{homes95}.

It is a challenge to theory to account for this simple conductivity behavior,
which is entirely unexpected. A simple anisotropic metallic system  would show
a Drude peak centered at zero frequency. If the observed spectrum is the low
frequency end of a {\it broad} Drude spectrum with an extremely short
relaxation time, a simple calculation would yield a mean free path of less than
a lattice spacing, which is in clear contradiction of the long mean free paths
seen in the ab-plane conductivity in samples from the same source\cite{bonn92}.
On the other hand, there is no sign of a {\it narrow} Drude peak that one might
expect if the system is an anisotropic metal with a very large mass in the
c-direction. At low frequency the observed conductivity is flat to less than 50
\cm\  and the lowest frequency conductivity is in excellent agreement with the
dc conductivity, which rules out any sharp Drude peak below the range of far
infrared spectroscopy\cite{homes95,homes95a}. However, it should be noted that
in overdoped samples a rather broad Drude peak appears, which is superimposed
on the broad back ground\cite{schutzmann94,bernhard98}.

There is a large literature that addresses the theoretical problem of this
incoherent c-axis
response.\cite{kumar92,chakravarty93,rojo93,clarke95,abrikosov96,atkinson99}.
Perhaps the most successful are the models that are founded on the original
suggestion of Anderson that in analogy with one-dimensional systems, the c axis
transport is between two Luttinger liquids. An expression for the conductivity
in the one dimensional case is given by Clarke et al. \cite{clarke95}
\begin{equation}
\sigma(\omega) \propto \omega^{4\alpha}
\end{equation}

where for the 1D Hubbard model $0<4\alpha<1/4$. This expression yields a very
flat conductivity if a value of $\alpha < 0.2$ is used.


%
%
\begin{figure}[t]
\vspace*{0.5cm}%
\centerline{\includegraphics[width=3.6 in]{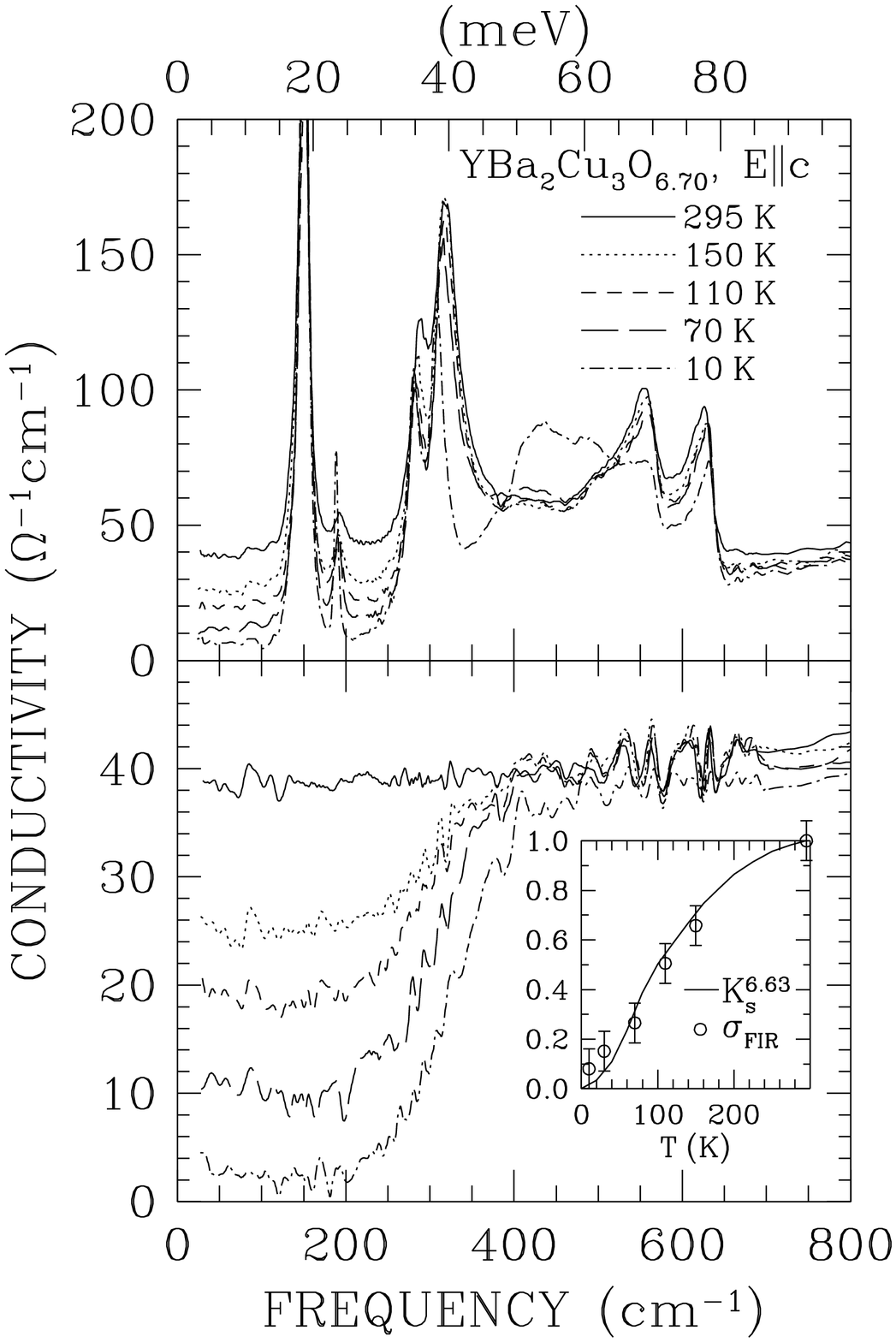}}%
\vspace*{0.0cm}%
\caption{Optical conductivity measured along the c-axis in \YBCO. The top panel
shows the measured conductivity showing prominent phonon peaks. In the bottom
panel these peaks have been subtracted. At high temperature the conductivity is
flat and frequency independent. As the temperature is lowered a clear pseudogap
is seen below 400 \cm, which appears well above the superconducting transition
temperature of 63 K. The amplitude of the gap (open circles) is compared with
the NMR
Knight shift in the inset.    }%
\label{lumpfig2}
\end{figure}

As the temperature is lowered a gap develops in the c-axis conductivity. To see
this clearly, the phonon lines have been fitted to Lorentzian profiles and
subtracted from the conductivity as shown in the lower panel of Fig. 2. In
addition to the phonon lines, the broad peak at 400 \cm\  that appears at low
temperature has also been subtracted. A clear gap-like depression of
conductivity is revealed below 400 \cm.  Several characteristic features of
this c-axis normal state gap should be noted. First, its frequency width does
not change with temperature. In other words, the gap does not close as the
temperature increases. Instead, as the temperature is raised the depth of the
gap decreases up to the point where, at 300 K, the gap has merged with the
background conductivity while the frequency scale remains fixed at
approximately 400 \cm. Secondly, it is {\it a normal state gap} and changes on
entry into the superconducting state at $T_c$ are relatively small. The third
striking property is the flat of the bottom of the gap, which remains flat to
high temperatures.

Optimally and overdoped YBCO also show a gap-like depression in low frequency
conductivity but the gap appears at $T_c$ and has been associated with the
superconducting gap\cite{koch90b,homes93,bernhard98}. This gap does not have
the flat bottomed shape of the underdoped materials. Instead, the conductivity
rises gradually from a low value at low frequency in a linear fashion to reach
the normal state value in the 600 \cm\  region.

In this connection it must be pointed out that the non-superconducting ladder
compound Sr$_{14-x}$Ca$_x$Cu$_{24}$O$_{41}$\cite{osafune99} where $x=11$, shows
a flat bottomed gap normal to the leg direction while the $x=8$ sample has a
gap similar to the optimally doped YBCO. Note that the $x=11$ material has the
higher doping and a higher conductivity and is considered to be akin to the
optimally doped YBCO. Thus in the ladder compounds the situation is reversed --
as the hole doping increases the low temperature gap becomes flatter. This
suggests that the shape of the gap at low temperature can not be used as a
signature of superconductivity in strongly correlated superconductors.

It has been suggested that the blocking layers play a role in the c-axis
transport\cite{abrikosov96,atkinson99}. In a series of experiments Bernhard et
al.\cite{bernhard98,bernhard00a} on calcium and Pr doped YBCO show that this is
not the case. Calcium and praseodymium have the effect of changing the hole
doping in the planes without affecting the oxygen content of the chains.
Calcium also allows a much larger overdoping of the copper-oxygen planes while
keeping the structure of the chains fixed as shown by the ratio of the
concentration of four-fold and two-fold coordinated copper chain copper sites.
This can be seen by the intensities of the phonons at the corresponding
bridging oxygen frequencies\cite{burns91,homes95a}. Praseodymium has the
opposite effect of reducing the in-plane doping. Bernhard et al. find that the
c axis spectra are independent of the amount of oxygen filling of the chains
and depend only on the doping of the planes ruling out models of c-axis
transport that involve details of the blocking layers. In the calcium doped
materials the authors do find that the overall conductivity in the c-axis
direction is higher, which they attribute to the effect of the chains.


On the other hand, substitution of Zn, a strong pair-breaking impurity, which
is known to reside on the copper oxygen planes has a strong effect on the
c-axis pseudogap spectra \cite{basov96c,bernhard00a}. Reducing $T_c$ with Zn
substitution is very different from what happens when $T_c$ is reduced by
underdoping. Basov et al. find that in naturally underdoped two-chain
YBa$_2$Cu$_4$O$_{8}$ (YBCO-124) zinc substitution at the 1.7 \% level has the
effect eliminating the c-axis pseudogap. This effect of zinc is confirmed by
Bernhard et al. who find, in optimally doped \YBCO\  that at zinc concentration
of 6 \% with $T_c$ of 64 K, the c-axis spectra show no pseudogap and resemble
overdoped samples with a broad Drude component superimposed on the flat c-axis
conductivity. In both systems, zinc doping also eliminates the broad peak at
400 \cm. We will return to the role of zinc substitution when we discuss the
effect of zinc on ab-plane spectra.


The peak in the conductivity spectrum at 400 \cm\  was originally interpreted
in terms of phonons\cite{homes93} based on the observation that showed that the
overall spectral weight was conserved in the phonon region between 250 \cm\ and
700 \cm\cite{homes95a}. In other words, the spectral weight of the peak at low
temperature grows at the expense of the spectral weight of some of the phonons.
However, more recently a much more satisfactory model has been proposed in
terms of an interlayer
plasmon\cite{vandermarel96,munzar99,bernhard00b,timusk03}. This model takes
into account the charge imbalance that is set up in a bilayer system where the
conductivity between the bilayers is much larger than between the charge
reservoir layers. The model not only explains the mode at 400 \cm\  but also
accounts in detail for the dramatic changes in intensity of certain oxygen
phonons\cite{munzar99}.

%
\begin{figure}[t]
\vspace*{2.5cm}%
\centerline{\includegraphics[width=3.5in]{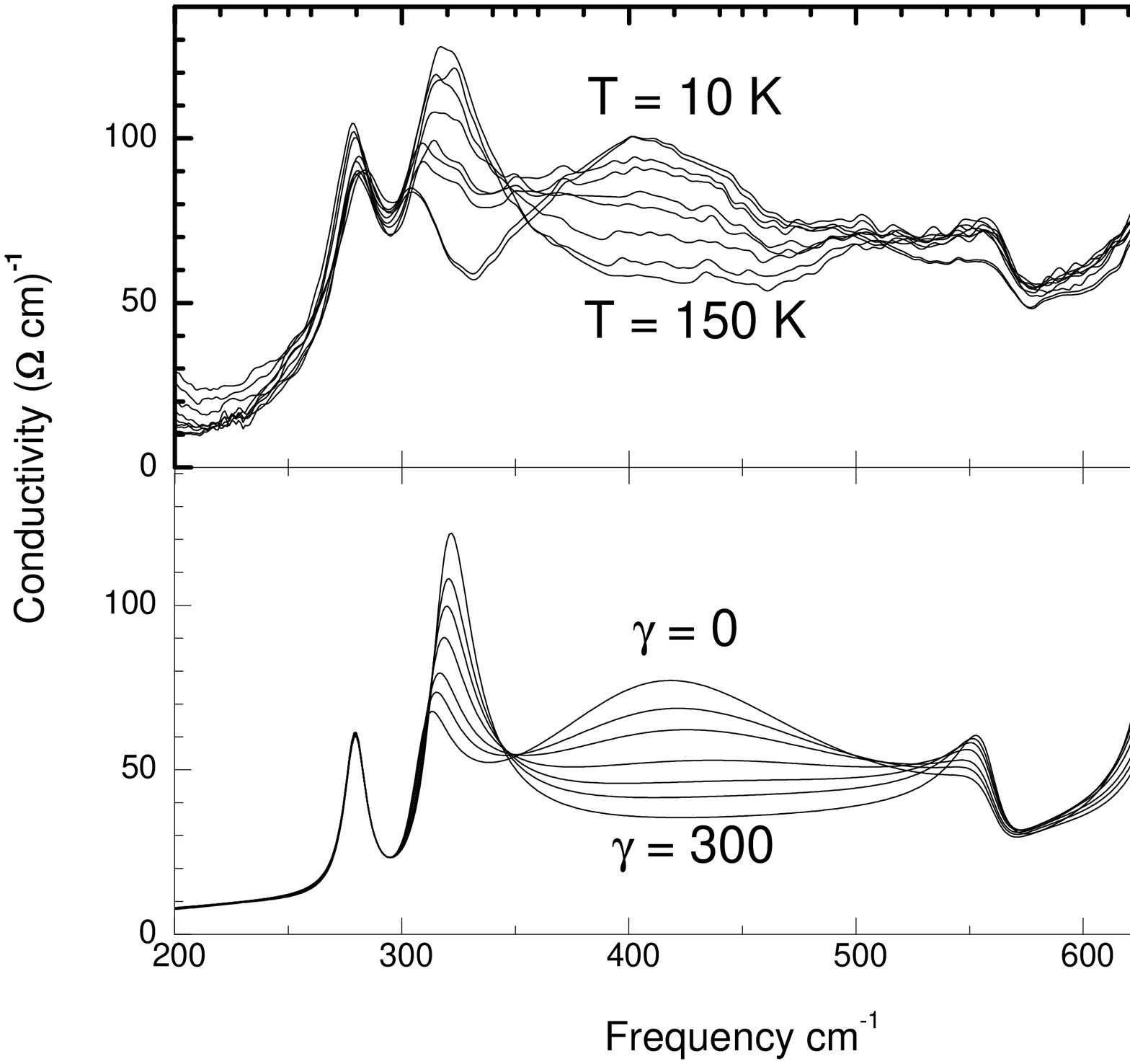}}%
\vspace*{-2.0cm}%
\caption{Interplane conductivity in the oxygen phonon region of underdoped
\YBCO, top panel. As the temperature is lowered below 150 K dramatic changes
occur in the intensities of the CuO$_2$ plane phonon frequencies. At the same
time a broad peak centered at 400 \cm\  grow in spectral weight. The lower
panel shows a calculated spectrum based on the interplane plasmon model of ref.
\cite{timusk03} where the damping rate of the interlayer plasmon
is allowed to vary .  }%
\label{isaksonfig3}
\end{figure}

Fig. 3 shows the temperature dependence of the 400 \cm\  mode in underdoped
YBCO, top panel, and a calculation based on the model of Munzar et al. in the
bottom panel\cite{timusk03}. It can be seen from the experimental data that the
mode makes its appearance below 150 K and thus is not associated with the
c-axis pseudogap, which at this doping level, starts to form at room
temperature. The calculation in the lower panel is based on the assumption that
the temperature dependence of the interlayer plasmon intensity is caused by
ab-plane life time effects\cite{timusk03}.

The frequency-width of the c-axis pseudogap is difficult to establish and there
is disagreement about its exact value and its temperature and doping
dependence. Homes et al. suggested a pseudogap in the conductivity of $\approx
400$ \cm, independent of temperature and doping in the doping range of 6.50 to
6.80, with a larger gap in the optimally doped materials. Bernhard, on the
other hand, argued for a larger gap based on c-axis measurements in underdoped
Ca doped samples\cite{bernhard99}. These discrepancies on gap size arise from
the difficulty in performing an accurate subtraction of the phonon background
in the 300 to 600 \cm\  region and to the question of whether there is a
depression of conductivity beyond the frequency of the highest frequency
phonons. Perhaps an analysis that uses detailed fits to the interlayer model of
Munzar et al.\cite{munzar99} can help to give a more reliable spectroscopy in
this frequency region.


More accurate data on the width of the pseudogap can be obtained from ARPES and
tunnelling measurements on underdoped Bi$_2$Sr$_2$CaCu$_2$O$_{8-\delta}$
(Bi-2212). Both techniques show a clear normal state gap, which persists nearly
to room temperature\cite{loeser96,ding96,renner98,krasnov00} A larger value of
the width of the pseudogap is found\cite{ding96,marshall96} with $2\Delta$ 480
to 800 \cm. In agreement with the infrared data, this gap does not close with
temperature but fills in, very much like the gap in the c-axis optical
conductivity. An important contribution of the ARPES technique is the finding
that the pseudogap develops in the ($\pi$,0) direction in the Brillouin zone
and has the same d-wave symmetry as the superconducting gap. The ARPES
observation that the normal state gap involves charge carriers away from the
nodes is in accord with band structure calculations that show that the matrix
element for interplane charge transfer is very small for momenta parallel to
($\pi$,$\pi$)\cite{andersen94,andersen94b,xiang96}. Thus the c-axis transport
involves carriers in the antinodal ($\pi$,0) region.

\section{The pseudogap and the ab plane scattering rate}

In many ways measurements of reflectance with infrared polarized in the
ab-plane are an ideal way to study low-lying excitations of high temperature
superconductors. One reason is a practical one: flux grown single crystals of
high temperature superconductors form as thin plates with a large ab-plane
area. As a result large ab-plane crystals have been grown for nearly all high
temperature superconductors. Also, unlike ARPES and tunnelling techniques,
optical techniques are not very surface sensitive.

The onset of absorption at 500 \cm\  for light polarized in the ab-plane, as
seen in Fig. 1, has been attributed to the superconducting gap or the
pseudogap. Like the pseudogap as it is seen in the c-axis conductivity, this
onset occurs in the normal state and the onset temperature decreases with
increased doping. However, it occurs at a much lower temperature than the
c-axis pseudogap in the same samples. Also, since the ab-plane transport is
coherent the gap cannot be a simple density-of-states effect.

%
\begin{figure}[t]
\vspace*{0.0 cm}%
\centerline{\includegraphics[width=4.0 in]{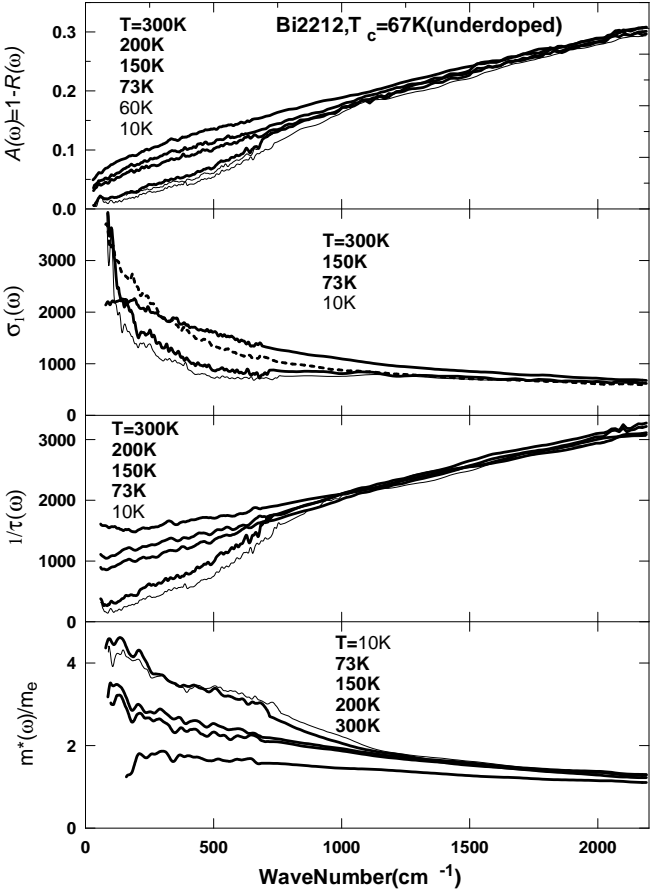}}%
\vspace*{1.0 cm}%
\caption{ab-plane transport properties of underdoped Bi-2212. The top panel
shows the absorbance, defined as $1-R$ where $R$ is the reflectance. In this
strongly underdoped sample an absorption threshold can be discerned at 600 \cm.
In the next panel the corresponding optical conductivity is shown. It is
approximately Drude-like without a clear gap.  The frequency dependent
scattering rate, shown in the next panel has a depression below 600 \cm. This
depression appears at temperatures below 150 K, well above the superconducting
transition temperature. The bottom panel shows the mass
enhancement.}%
\label{isaksonfig4}
\end{figure}

The relationship between a gap in the density of states and an onset of
infrared absorption is indirect. The most straightforward way to show this
relationship is to first perform Kramers-Kronig analysis on the reflectance to
obtain the real and imaginary parts of the conductivity. An example of this
kind of analysis is shown in Fig. 4. The absorbance $A$ of an underdoped sample
of Bi-2212 with a $T_c$ of 67 K and the optical conductivity are shown from the
work of Puchkov et al.\cite{puchkov96c} While the reflectance shows a clear
onset of absorption in the 600 \cm\  region, there is no gap in the
conductivity. The conductivity looks Drude-like with only a hint of a peak with
an onset in the spectral region of the absorption threshold in reflectance. To
clearly reveal the presence of any gap it is necessary to look at the frequency
dependent scattering rate using the extended Drude model of Allen and
Mikkelsen.\cite{allen77,basov96c}


Based on earlier theoretical work\cite{mori65,gotze72} Allen and Mikkelsen
showed that the Drude formula

\begin{equation}
\sigma(\omega)= {1 \over {4\pi}} {{\omega_p^2} \over {1/\tau-i\omega}}
\end{equation}

can be extended to include a frequency dependent scattering rate:

\begin{equation}
\sigma(\omega,T)= {1 \over {4\pi}}{{\omega_p^2} \over {1/\tau(\omega,T)-i\omega[1+\lambda(\omega,T)]}}
\end{equation}
where $1/\tau(\omega,T)$ is the frequency--dependent scattering rate and
$\lambda(\omega,T)$ the mass enhancement of the electronic excitations due to
the interactions.

One can solve for $1/\tau(\omega)$ and $1+\lambda(\omega)$ in terms of the
optical conductivity found from the experimental reflectivity to find:

\begin{equation}
1/\tau(\omega)={\omega_p^2 \over {4\pi}} Re({1 \over \sigma(\omega)}),
\label{tausig}
\end{equation}
The dc resistivity is zero frequency limit
$\rho_{dc}(T)=1/\sigma_{dc}(T)=m_e/(\tau(T)ne^2)$ since $\sigma(\omega)$ is real in the
zero frequency limit.

The mass enhancement factor $\lambda(\omega)$ is the imaginary part of
$1/\sigma(\omega)$:

\begin{equation}
1+\lambda(\omega)=-{\omega_p^2 \over {4\pi}} {1 \over \omega} Im({1 \over
\sigma(\omega)}).
\end{equation}
The total plasma frequency $\omega_p^2$ can be found from the sum rule
$\int_0^{\infty}\sigma_1(\omega)d\omega=\omega_p^2/8$. The choice of this
arbitrary cut-off introduces an uncertainty to the overall scale factor of the
scattering rate and effective mass.

The frequency dependent scattering rate formalism has been used to describe
electron--phonon scattering\cite{allen71,shulga91}. Shulga et
al.\cite{shulga91} give the following expression for $1/\tau(\omega,T)$:

\begin{eqnarray}
{1 \over \tau}(\omega,T)={\pi \over \omega} \int_0^{\infty} d\Omega
\alpha_{tr}^2(\Omega)F(\Omega) \big[2{\omega}{\coth}({\Omega \over {2T}})- \\
-(\omega+\Omega){\coth}({{\omega+\Omega} \over {2T}})+ \nonumber
(\omega-\Omega){\coth}({{\omega-\Omega} \over {2T}})\big] + {1 \over
\tau_{imp}}. \label{Shulga}
\end{eqnarray}
Here $\alpha_{tr}^2(\Omega)F(\Omega)$ is a weighted phonon density of states
and $T$ is the temperature measured in frequency units. The last term
represents impurity scattering.

The quantity $\alpha_{tr}^2(\Omega)F(\Omega)$ is closely related to
$\alpha^2(\Omega)F(\Omega)$ obtained from the inversion of tunnelling spectra
in BCS superconductors\cite{allen71} and it can found by inverting the optical
conductivity spectra at $T=0$\cite{allen71,farnworth74,marsiglio98} to yield:

\begin{equation}
\alpha^2(\Omega)F(\Omega) = {1 \over 2\pi\omega}{\partial \over
\partial\omega}\Bigg[\omega^2{\partial \over \partial\omega}{1 \over
\tau(\omega)}\Bigg]. \label{a2F}
\end{equation}
The difficulty with the application of this method to practical spectroscopy
arises from the presence of the second derivatives in the expression, which
places severe demands on the signal-to-noise ratio of the experimental spectra.
Despite this, the method has been applied with considerable success to BCS
superconductors to extract $\alpha^2(\Omega)F(\Omega)$, to
K$_3$C$_{60}$\cite{marsiglio98} and more recently to high $T_c$
superconductors\cite{carbotte99,singley01,tu01}.

There is an alternative approach to the analysis of infrared data, the
two-component model\cite{tanner92,gao93}. Here one fits a Drude oscillator to
the low frequency component of the conductivity and treats the remaining
conductivity as a separate parallel channel of conductivity, the {\it
mid-infrared} band. This approach is applicable in particular to the La-214
system where such a separate band is more clearly resolved but it has also been
used for the cases where there is no obvious separation between the low and
high frequency portions of the spectrum\cite{quijada99}.


Fig. 4, third and fourth panels from the top, show the scattering rate and the
effective mass for the underdoped Bi-2212 system. We see that the scattering
rate at room temperature has a linear variation with frequency starting from a
rather large constant value 1280 \cm. The scattering rate above 1000 \cm\
remains temperature independent at its 300 K value but as the temperature is
lowered there is a lowering of low frequency scattering between $T^*$ and
$T_s$. At 150 K the scattering rate begins to drop precipitously at low
frequency forming a gap-like depression that extends to about 700 \cm. To
distinguish its onset temperature from the higher onset temperature of the
c-axis pseudogap $T^*$, we will call the lower ab-plane onset
$T_s$\cite{timusk03}. Like the c-axis pseudogap it occurs in the normal state
in underdoped samples and at or near $T_c$ in optimally doped
ones\cite{puchkov96d}. In underdoped systems the change in the scattering rate
on entering the superconducting state is small. The lowest panel shows the mass
enhancement of the carriers resulting from the frequency dependent scattering.

These general features are seen in all the cuprate systems. A gap in the
scattering rate was first identified in YBCO and Bi-2212 systems, materials
with two copper-oxygen layers, but it is also seen in the one-layer
Tl$_2$Ba$_2$CuO$_{6+\delta}$ (Tl-2201) system.\cite{puchkov96c} The depression
of low frequency scattering has also been reported in
La$_{2-x}$Sr$_x$Cu$_3$O$_{4}$ (La-214)\cite{startseva99u,startseva99o,dumm02}
and the three layer Hg$_2$Ba$_2$CaCu$_3$O$_{8+\delta}$ (Hg-2213)
systems\cite{mcguire00} as well as in the electron doped
Nd$_{2-x}$Ce$_x$Cu$_3$O$_{4+\delta}$ (NdCe)\cite{singley01,dumm02}.

The three layer mercury compound  (Hg-1223) is unique in that it has a maximum
$T_c$ of as high as 135 K, which is substantially higher than the intensively
studied "canonical" superconductors YBCO and Bi-2212 both with maximum $T_c$'s
at optimal doping of $\approx$93 K. Single crystals of the mercury material are
difficult to grow and, and as a result, very few experiments have been done on
this system. Infrared spectroscopy on the ab-plane that has been performed by
McGuire et al.\cite{mcguire00} shows that the material behaves very much like
other cuprates with one notable exception: the energy scale for the gap in
ab-plane scattering is 40 \% higher than in materials with a $T_c$ of 93 K, in
the exact ratio of their transition temperatures. Energy scales measured with
other probes such as the B$_{1g}$ Raman intensity\cite{sacuto98} and the copper
hyperfine coupling constant measured by NMR\cite{julien96} are also much
higher.

The La-241 material also differs from YBCO and Bi-2212 in that its $T_c$ at
optimal doping is substantially {\it lower} than YBCO and Bi-2212. The presence
of a pseudogap in this material is somewhat controversial. Magnetic resonance
suggests that the pseudogap is unusually weak\cite{millis93} and the copper
spin relaxation time shows no sign of a pseudogap\cite{ohsugi91}. On the other
hand, transport measurements find a suppression of resistivity at a temperature
$T^*$, which is unusually high\cite{batlogg94}. Infrared conductivity in the
c-axis direction shows the presence of a pseudogap in the conductivity that
starts near room temperature in slightly underdoped samples, but the overall
suppression of conductivity is less complete than in the YBCO
material\cite{basov95c,uchida96,startseva99u}.

In the ab-plane infrared properties La-241 also shows differences when compared
to other cuprates. From its lower superconducting transition temperature one
might expect a overall lower frequency scales. This is not the case. The
characteristic knee in reflectance that signals the onset of scattering is seen
in the La-214 material at approximately the same frequency range as in
materials with a $T_c$ of 93 K\cite{startseva99u,startseva99o}. The exact
frequency of this feature is difficult to discern due to the presence of c-axis
longitudinal phonon lines induced by structural defects\cite{reedyk92,tajima}.
Recent experiments with better crystals\cite{dumm02} suggest the gap in the
scattering rate in La-214 is around 1000 \cm, similar to that in
superconductors with higher transition temperatures but it is difficult to
separate this energy scale from the lower scale caused by the resonance as
shown in Fig. 4 for Bi-2212.

Because the ab-plane conductivity is metallic and approximately Drude-like, the
gap-like depression seen in the scattering rate cannot simply be interpreted as
a gap in the density of states. As the frequency dependent scattering rate
formalism suggests, we must look to models that describe the scattering rate.
The most successful of these have been the marginal Fermi liquid model (MFL) of
Varma et al.\cite{varma89} at high temperatures and the scattering of carriers
by a bosonic excitation\cite{thomas88,norman99,carbotte99,abanov01} at low
temperatures. It must be pointed out however that the models that invoke a
bosonic excitation to account for the onset of scattering at $T_s$ are based on
Fermi liquid ideas with well defined quasiparticles. As an inspection of the
scattering rates shown in Fig. 4 shows that for this material at least the
scattering rate is higher than the frequency at all temperatures.


As Fig. 4 shows, the scattering rate in underdoped Bi-2212 at \emph{room
temperature} is quite remarkable with a frequency variation that is accurately
linear up to at least 2000 \cm. It is worth pointing out that a Drude model
predicts a constant scattering rate as a function of frequency and any model
involving bosonic excitation would have characteristic features at the
frequency of these excitations. This simple featureless linear behavior can be
seen in all cuprate materials and at all doping levels and is closely related
to another remarkable property of the cuprates, their linear variation of
resistivity with temperature\cite{gurvitch87}. The semi-empirical MFL model of
Varma et al.\cite{varma89} attempts to account for these properties by assuming
that the electron self energy $\Sigma({\bf k},\omega)$ rate is given by:
\begin{equation}
\Sigma({\bf k}) \approx g^2N^2(0)\big(\omega\ln{x \over \omega_c} - i {\pi
\over 2}x\big)
\end{equation}
where $g$ is a coupling constant, $N(0)$ the density of states at the Fermi
surface, $\omega_c$ a high frequency cut-off and $x=\max(|\omega|,T)$. This
expression fits the experimental data well in the high temperature
region\cite{varma89,schlesinger90} but fails in the underdoped region at low
temperatures since it does not include the gap-like depression in scattering
below the $T_s$ temperature.


It is clear from Fig. 4 that, as the temperature is lowered, a gap develops in
the scattering rate below 300 K \cite{Ts temperature}. The gap has the effect
of reducing the scattering below the marginal Fermi liquid line. In optimally
doped samples of YBCO this temperature is very close to the superconducting
transition temperature\cite{basov96c}, while in Bi-2212 it is clearly above
$T_c$\cite{puchkov96c,tu01}. However, while a gap in scattering may be caused
by a gap in the density of states, as for example in the case of simple
electron-electron scattering, it is now clear that this model does not account
all the experimental data in underdoped cuprates. As the data in Fig. 4 show,
the gap in the scattering rate in underdoped Bi-2212 first sets in the around
room temperature with a frequency scale of the order of 1000 \cm\ followed by a
more rapid decrease in neighborhood of 150 K. The lower temperature gap can be
understood, in analogy with the electron-phonon interaction in conventional
superconductors\cite{allen71}, where the scattering by a bosonic mode is
invoked to account for the onset of scattering in the 500 \cm\ region.

An early attempt along these lines was the work of Thomas et al.\cite{thomas88}
and Puchkov et al.\cite{puchkov96d} who found that a broad bosonic peak in the
300 \cm\  region could account for the scattering rate variation with frequency
although not with temperature. More recently, Marsiglio showed that one could
extract the spectral density of this mode $W(\omega)$ from the experimental
data by taking the second derivative of the scattering rate\cite{marsiglio98}:
\begin{equation}
W(\omega) = {1 \over 2\pi}{d^2 \over d\omega^2}{\omega \over \tau(\omega)}.
\end{equation}
This method has been successfully applied to the scattering rate spectra of
C$^{60}$\cite{marsiglio98}, to YBCO\cite{carbotte99,schachinger01}. This
analysis did not account too well for a region of negative $W(\omega)$ seen
above the broad peak in the superconducting state.

Abanov et al.\cite{abanov01} showed that the negative peak was not only
expected but gave an independent measure of the superconducting gap $\Delta$.
Using the method of Abanov, Tu et al. analyzed very high quality reflectance
data on optimally doped Bi-2212\cite{tu01}  and were able to deduce a value of
the superconducting gap $\Delta=30 \pm 4$ meV and a resonance $\omega_sr=40 \pm
4$ meV (322 \cm) in excellent agreement with tunnelling data at $\Delta=27$ meV
and the mode seen in magnetic neutron scattering $\omega_{sr}=41$
meV\cite{bourges98}. In the normal state, the peak is seen at a frequency
corresponding to the resonance mode and the peak persists at least up to 100K
and the negative going feature due to superconductivity can not be seen. These
results are in accord with the observation that in optimally doped \YBCO\ where
the infrared mode is not seen in the normal state and turns on sharply at
$T_c$\cite{puchkov96c}. However, we note that while in optimally doped Bi-2212
a clear kink was seen in the scattering rate at $T_c$ whereas the YBCO spectrum
was flat and featureless at 95 K, a few degrees above $T_c$. This suggests that
in optimally doped \YBCO the mode-scattering onset temperature $T_s \approx
T_c$, while in Bi-2212 $T_s > T_c$.

%
\begin{figure}[t]
\vspace*{2.5cm}%
\centerline{\includegraphics[width=3.5in]{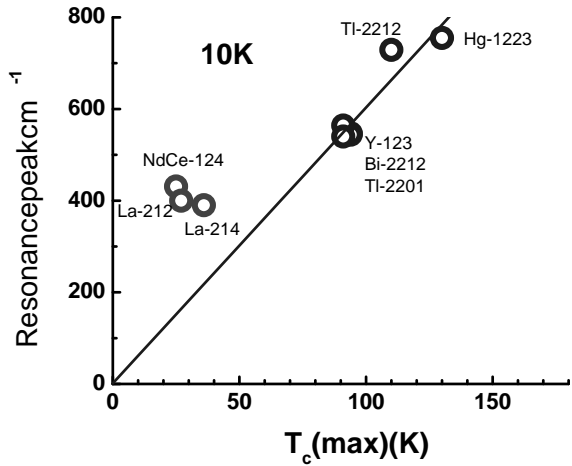}}%
\vspace*{-2.0cm}%
\caption{The frequency of the ab-plane scattering resonance as a function of
$T_c$ for optimally doped high $T_c$ supercondcutors. The straight line is a
fit to the superconductors with a $T_c$ above 90 K. Included in this group is
the one-plane Tl-2201. The LaSr-214 and NdCe-214 do not follow on this line. }%
\label{isaksonfig5}
\end{figure}

To further investigate the systematics of the bosonic mode, we plot in Fig. 5
the frequency of the peak in $W(\omega)$  at low temperature as a function of
$T_c$ for several optimally doped cuprate systems. Most of the data are from a
recent paper by Singley et al.\cite{singley01} with added data on Tl-2212 from
the work of Tanner\cite{tanner private}. Two features stand out. First, for
most high $T_c$ systems the points cluster at $T_c= 93$ K and peak frequency of
550 \cm. However, if the data from the mercury and the two layer thallium
materials are included, there is a trend for the systems with higher $T_c$
towards a linear dependence with a regression line that goes trough the origin.
Within the bosonic model, this means that at optimal doping the sum of the
superconducting gap and the resonance frequency is proportional to $T_c$.

The second notable feature of Fig. 5 is that the linear variation does not
apply to the LaSr and NdCe systems. While the systems do exhibit a peak at
reasonably high frequency, their $T_c$ seems to be suppressed as compared to
the systems with higher transition temperatures. We note that this is not an
effect of double vs. single layers since one of the systems that falls on the
line is the one-layer Tl-2201 system. The $T_c$ suppression in the LaSr system
may be due to the disorder introduced by the La dopant ions. But it is also
possible that the model of scattering by a well defined mode is not appropriate
for these systems and that the gap-like depression of scattering is due to the
combination of a pseudogap and a broad mode. In this context it is worth
pointing out that the neutron mode has not been seen in these systems.

The bosonic mode model is expected to give rise to a kink in the dispersion
curves of the charge carriers\cite{norman99,eschrig00} and this kink has indeed
been observed in ARPES dispersion curves\cite{kaminski00,bogdanov00}, which
further confirms the overall picture of the kink. The frequency of the kink in
ARPES dispersion is in fairly good agreement with the frequency of the bosonic
mode deduced from optical spectra\cite{norman99} and its temperature dependence
is in general agreement with what is seen in optical spectra.

While the picture of the bosonic mode explains a number of features in infrared
and ARPES spectra, it does not provide a complete theory of the interactions of
the holes in high temperature superconductors since the model tells us nothing
about the nature of the mode. Carbotte et al.\cite{carbotte99} proposed that
the 41 meV neutron mode is the responsible agent.  Lanzara et al. on the other
hand, have made the suggestion recently that the mode is due to an in-plane
copper-oxygen vibration.\cite{lanzara01} We will examine these ideas in turn.

The picture of the neutron mode as the cause of the gap-like depression of the
ab-plane scattering rate below a temperature $T_s$ and the kink in the ARPES
dispersion spectra has several attractive features. The frequency of the
neutron peak is in good agreement with the observed position of the infrared
features\cite{carbotte99,tu01} in systems where the neutron peak has been seen.
Unfortunately this is not a good test since, as fig. 5 shows, all these systems
have the same frequency scale. The crucial test, the observation of the neutron
peak in the three layer mercury compound, is very difficult because of the lack
of large enough crystals for magnetic neutron scattering.

Secondly, the temperature where the neutron peak appears is in accord with the
onset temperature $T_s$ of depressed ab-plane scattering\cite{puchkov96d}
although there is an unfortunate lack of data in the crucial temperature region
between 200 K and $T_c$ where the gap in the scattering rate makes its
appearance. It is clear that more work is needed to settle this question. ARPES
spectra also show that the kink and the accompanying decrease in scattering at
low temperature are consistent with the neutron mode picture but the ARPES data
are also very sparse along the temperature axis.

The alternate picture of a phononic origin of the mode is more difficult to
accept. First of all, the mode frequency does not show the isotope effect
expected for phonons\cite{wang02}. Secondly, since the mode frequency scales
with $T_c$ as shown in Fig. 5, which is the sum of the gap frequency and the
mode frequency, it is difficult to understand the 40 \% increase in this sum in
the three layer mercury compound in the phonon picture since copper-oxygen
in-plane phonon frequencies do not change substantially from one material to
another. Also, quantitatively,  phonon models based on a Fermi liquid approach
fail completely when it comes to the temperature dependence of the scattering
rate\cite{puchkov96d,singley01}. The neutron mode, unlike the phonons, has a
strongly temperature dependent spectral function, which essentially vanishes at
a temperature not far above $T_s$ into the general magnetic background.

Further support for a magnetic origin of the mode is offered by Zn doping
experiments. Recent experiments\cite{fong99} show that Zn doping has the effect
of reducing the width of the neutron mode while leaving the overall spectral
weight unchanged.  This is in accord with the general picture of zinc as a
strong in-plane scattering center as seen from the strong effect of zinc on
both $T_c$ and resistivity. The ab-plane scattering rate spectra of Basov et
al.\cite{basov96c} confirm this picture for the naturally underdoped system of
YBCO-124. In the undoped material the scattering rate increases sharply
$\approx 350$ \cm\  followed by a broader threshold at 500 \cm,  which is very
similar to what happens in other underdoped materials.  With 0.425 \% zinc
doping the sharp onset structure is completely obliterated but an overall
depression of low frequency scattering remains and is spread out over a
frequency interval of the order of 1000 \cm. This doping level is evidently
high enough to suppress the scattering by a well defined mode but retain the
influence of the pseudogap in the density of states. Further doping to 1.275 \%
zinc has the interesting effect of completely suppressing superconductivity and
eliminating any gap-like structure in the scattering rate spectra, which makes
them look very much like the spectra in the normal state of optimally doped
YBCO where the neutron mode is absent.

Similar effects are found by Zheng et al. in NMR experiments on YBCO-124 doped
with zinc\cite{zheng96,zheng93}.  While Zn doping has a relatively small effect
on the Knight shift, it completely suppresses the pseudogap in the spin
relaxation rate already at the 1\% doping level.

\section{Evidence for two characteristic energy scales.}

All experimental techniques used to study the cuprates show evidence of a
pseudogap in underdoped materials\cite{timusk99}.  Common features include a
decreasing onset temperature $T^*$ with doping, a temperature that appears to
merge with the superconducting transition temperature at or near optimal
doping. However, early on there was evidence from NMR experiments for {\it two}
energy scales, a higher one seen in the $q=0$ susceptibility as revealed by the
Knight shift, and a lower one related to the $\pi,\pi$ susceptibility of the
relaxation rate $1/T_1T$\cite{nmr gaps}. Generally the Knight shift is
considered to be a measure of the density of states whereas the relaxation rate
measures the carrier life time.

That the gap in the c-axis optical conductivity and the NMR Knight shift may
have a common origin can be seen from the temperature scale of the Knight shift
and the optical conductivity shown as the inset of Fig. 2 where both extend to
almost room temperature in the underdoped YBCO with $T_c= 60 $ K. A similar
correspondence has been found for the double chain compound
YBCO-124\cite{basov94c}. The broad suppression of frequency dependent
scattering with the 1000 \cm\  frequency scale seen in Bi-2212 must also be
included in this class of phenomena.

However, there is good evidence for another class of phenomena with a lowere
temperature and frequency scale. In the underdoped YBCO $T_c= 60 $ K material,
the onset of several physical quantities relating to the life time of the
in-plane excitations occur at a much lower temperature, in the 150 K region,
and they have their most rapid change at $T_c$. This is illustrated in Figs. 6
and 7 from a recent paper of Timusk and Homes\cite{timusk03} where in addition
to the NMR Knight shift and the relaxation $1/T_1T$, the c-axis pseudogap
amplitude and the amplitude of the c-axis infrared mode at 400 \cm\ have been
plotted.

%
\begin{figure}[t]
\vspace*{2.5cm}%
\centerline{\includegraphics[width=3.5in]{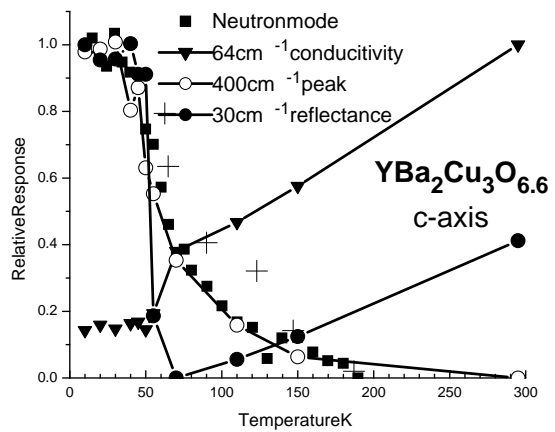}}%
\vspace*{-2.5cm}%
\caption{  Three temperature scales can be identified in underdoped \YBCO.  The
largest is around 300~K and is associated with the pseudogap, which is
approximated as the conductivity at 64~cm$^{-1}$ in c-axis transport.  The next
governs the intensity of the 400~cm$^{-1}$ peak and neutron mode. It has a
value of approximately 150~K as well as NMR relaxation, not shown here. The
lowest one is associated with superconductivity and has a value of 60~K in this
underdoped sample. We use the 30~cm$^{-1}$ reflectance to approximate the
superconducting condensate density. The crosses show an estimate of the 400
cm$^{-1}$ peak
amplitude based on broadening by in-plane scattering as shown in Fig. 3.}%
\label{isaksonfig6}
\end{figure}

It can be seen from Fig. 6 that the various plotted quantities fall into two
groups: first there are those that have an onset near room temperature and show
a rather gradual development of the amplitude as the temperature is lowered
with no rapid changes at the superconducting transition temperature. They
include the c-axis pseudogap and the NMR Knight shift. Not plotted here is the
tunnelling density of states, which also shows a pseudogap that also follows
the higher energy scale\cite{renner98,krasnov00}.

Then there is a second group of quantities that start at 150 K and change
rapidly near the superconducting transition temperature. These include the
relaxation $1/T_1T$, the amplitude of the c-axis peak at 400 \cm\ and the 41
meV neutron resonance.

At this stage it is not possible at this point to conclusively fit the ab-plane
scattering rate data into this picture. From the existing data it appears that
suppression of ab-plane scattering is caused by both the high temperature
phenomena in the form of the broad depression of scattering with the 1000 \cm\
scale and the low temperature gap in scattering associated with the bosonic
mode.  There is lack of relevant data not only for the infrared scattering rate
but also for the kink in ARPES dispersion and dc transport. More work is needed
to settle this question. However there are strong hints from the available data
that the ARPES kink and the ab-plane scattering rate gap from the bosonic mode
both belong to the low temperature group while the ARPES d-wave gap at
($\pi,\pi$)\cite{harris96} and the Knight shift along with with the 1000 \cm\
depression of conductivity follow the high temperature scale.

%
\begin{figure}[t]
\vspace*{2.5cm}%
\centerline{\includegraphics[width=3.5in]{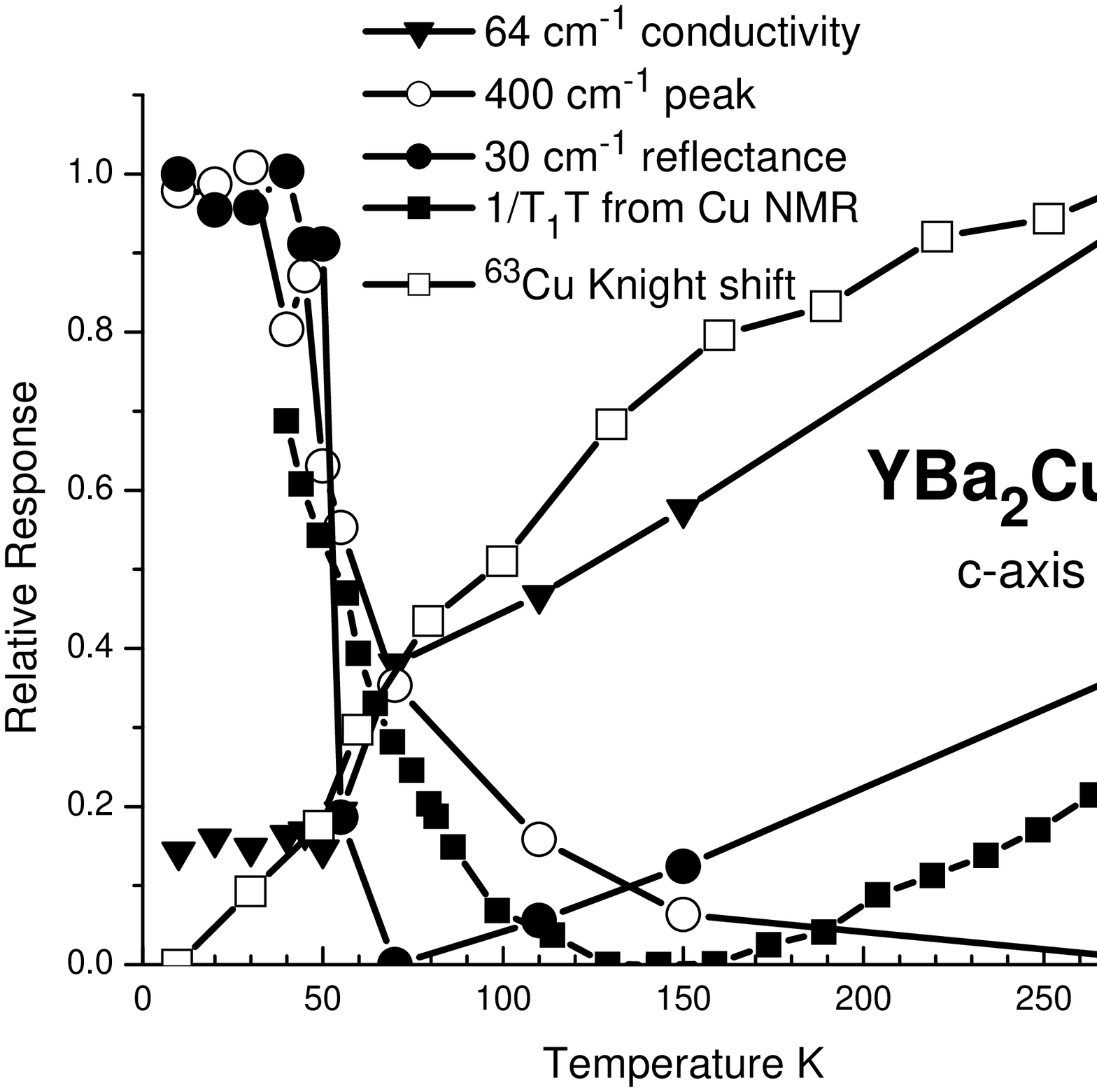}}%
\vspace*{-2.5cm}%
\caption{Infrared pseudogap and magnetism in underdoped YBCO.  The amplitude of
the Knight shift plotted as open squares and the pseudogap plotted as triangles
and approximated here as the 64~cm$^{-1}$ conductivity.  The Knight shift
depression starts at room temperature in this sample with an oxygen content of
6.60.  The relaxation time, shown as solid squares, is plotted as
$[1-1/(T_1T)]$.  It starts to deviate from the uniform high temperature form at
around 150~K and fits better to the 400~cm$^{-1}$ peak intensity than the
Knight shift.  }%
\label{isaksonfig7}
\end{figure}

\section{Summary}

We have discussed two separate experiments where infrared spectroscopy shows
evidence of a gap-like depression in physical quantities in underdoped
cuprates, the c-axis conductivity and the ab-plane scattering rate. The
depression of the c-axis conductivity seems to be associated with a real gap in
the density of states as shown by the Knight shift, ARPES and tunnelling and
possibly the 1000 \cm\  scale ab-plane scattering rate suppression. This
pseudogap appears near room temperature in samples where $T_c$ has been reduced
to 60 K, has width that does not change with temperature and seems to merge
with the superconducting gap near optimal doping. The gap associated with a
depressed scattering rate in the ab-plane transport with the 600 \cm\
frequency scale shows up at a lower temperature and appears to be associated
with scattering from a mode. ARPES dispersion shows this mode to be present in
Bi-2212 and there is good evidence that the 41 meV neutron mode is the
responsible agent, although the idea that phonons may be involved has also been
advanced. To confirm this picture of a gap and a distinct mode, a detailed
study of the temperature dependence of the two phenomena in the same system is
needed. Underdoped YBCO seems to be a good candidate for this since at least
two groups are able to grow thick, high-quality crystals of this
material\cite{erb_liang}.

\section{Acknowledgements}

First and foremost I would like to thank my long term collaborators who have
made this work possible. These include Doug Bonn, David Tanner, Maureen Reedyk,
John Greedan, Chris Homes, Dimitri Basov and Anton Puchkov.  More recently
Tatiana Startseva, Jeff McGuire, Nan Lin Wang, Toomas R\~{o}\~{o}m and Jungseek
Hwang have made enormous  contributions by developing the capability to work
with more and more difficult samples. My colleagues Elihu Abrahams, Phil
Anderson, John Berlinsky, Jules Carbotte, Takashi Imai, Steve Kivelson,
Catherine Kallin, Patrick Lee, Kathy Levin, Frank Marsiglio, Thom Mason, Andy
Millis, Mike Norman, David Pines, Louis Taillefer and Mike Walker have provided
valuable insight into the mysteries of the pseudogap in high temperature
superconductors.


\begin{thebibliography}{99}


\bibitem{tinkham} R.E. Glover and M. Tinkham, Phys. Rev B
{\bf 107}, 844, 1956;{\it ibid}, {\bf 108}, 1175 (1957).

\bibitem{joyce70} R.R. Joyce and P.L. Richards, Phys. Rev. Lett. {\bf 24},
1007 (1970).

\bibitem{farnworth74} B. Farnworth and T. Timusk, Phys. Rev B
{\bf 10}, 2799, (1974).

\bibitem{farnworth76} B. Farnworth and T. Timusk, Phys. Rev B {\bf 14},
5119 (1976).

\bibitem{brandli72} G. Br\"andli and A.J. Sievers, Phys. Rev B {\bf 5},
3550 (1972).

\bibitem{mattis58} D.C.~Mattis and J.~Bardeen, Phys. Rev. {\bf
111,} 412 (1958).

\bibitem{allen71} P.B. Allen, Phys. Rev B {\bf 3}, 305 (1971).
\comment{electron phonon interaction in optical spectra, relaxation region}

\bibitem{nam67} S.B.~Nam, Phys. Rev. {\bf 156}, 470, 487 (1967).

\bibitem{marsiglio98} F. Marsiglio, T. Startseva, J.P.~Carbotte,
Phys. Lett. A \textbf{245}, 354 (1998).


\bibitem{timusk88c} T. Timusk, S.L.~Herr, K.~Kamar\'as, C.D.~Porter, D.B.~Tanner,
D.A. Bonn, J.D.~Garrett, C.V.~Stager, J.E.~Greedan, and M.~Reedyk, Phys. Rev. B
{\bf 38}, 6683 (1988). \comment{BCS analysis of clean superconductors}

\bibitem{kamaras90} K. Kamar\'as, S.L. Herr, C.D. Porter, N.
Tache, D.B.Tanner, S. Etemad, T. Venkatesan, E. Chase, A. Inam, X.D. Wu, M.S.
Hegde,and B. Dutta, Phys. Rev. Lett. {\bf 64}, 84 (1990). \comment{``In a Clean
High-Tc SuperconductorYou Do Not See the Gap''}

\bibitem{reedyk88}  M.~Reedyk, D.A. Bonn, J.D.~Garrett, J.E.~Greedan, C.V.~Stager,
T. Timusk, K.~Kamar\'as, and D.B.~Tanner, Phys. Rev. B {\bf 38}, 11981 (1988).
\comment{\BiSr xtals, reflectance first observation of knee in normal state}

\bibitem{thomas88} G.A. Thomas, J. Orenstein, D.H. Rapkine, M.
Capizzi,A.J. Millis, R.N. Bhatt, L.F. Schneemeyer, and J.V. Waszczak, Phys.
Rev. Lett. {\bf 61}, 1313 (1988) \comment{``Ba2YCu3O7-d: Electrodynamics of
Crystals with HighReflectivity''}

\bibitem{basov96c} D.N.~Basov, R.~Liang, B.~Dabrowski, D.A.~Bonn,
W.N.~Hardy,and T.~Timusk,  Phys. Rev. Lett. {\bf 77}, 4090 (1996). \comment{
\YBa Pseudogap and charge dynamics in CuO$_2$ planes in YBCO.  28 ab plane
c-axis \YBa 123 124 and 124:Zn compared.}

\bibitem{carbotte99} J.P.~Carbotte, E.~Schachinger, and D.N.~Basov, Nature
\textbf{401}, 354 (1999). \comment{Coupling strength of charge carriers to spin
fluctuations in high-temperature superconductors. }

\bibitem{abanov01} A. Abanov, A.V. Cubukov, and J. Schmalian,
Phys. Rev B \textbf{63}, 180510 (2001).

\bibitem{bonn87} D.A. Bonn, J.E. Greedan, C.V. Stager, T. Timusk,
M.G.Doss, S.L. Herr, K. Kamar\'as, C.D. Porter, D.B. Tanner, J.M. Tarascon,
W.R.McKinnon, and L.H. Greene, Phys. Rev B {\bf 35}, 8843, (1987).

\bibitem{tamasaku92} K. Tamasaku, Y. Nakamura, and S. Uchida, Phys.
Rev. Lett. {\bf 69}, 1455 (1992) \comment{ \LaSr c-axis single crystal 20-350
cm-1 8K - 300 K  Charge dynamics across, Josephson plasma resonance...}

\bibitem{bonn92} D.A. Bonn, P. Dosanjh, R. Liang, and W.N. Hardy,
Phys. Rev. Lett. {\bf 68}, 2390 (1992). \comment{quasiparticle lifetime loss at
Tc \YBa microwave surface impedance}

\bibitem{basov01} D.N.~Basov, and T.~Timusk, Infrared Properties of High-$T_c$
Superconductors: an Experimental Overview in {\it Handbook on the Physics and
Chemistry of Rare Earths}, Vol 31, Edited by K.A. Gschneidner, Jr. L. Eyring
and M.B. Maple,  Elsevier Science BV, 2001.pp. 437-507, (2001).

\bibitem{timusk99} T. Timusk and B. Statt,  Reports of Progress in Physics {\bf
62}, 61 (1999).

\bibitem{warren89} W.W.~Warren, Jr., R.E.~Walstedt, G.F. Brennert,
R.J.~Cava, R.~Tycko, R.F.~Bell, and G.~Dabbagh, Phys. Rev. Lett. {\bf 62} 1193,
(1989). \comment{first mention of a spin gap.}

\bibitem{alloul89} H. Alloul, T. Ohno, and P. Mendels, Phys. Rev.
Lett. {\bf 63}, 1700 (1989). \comment{pseudogap NMR Y nuclei}

\bibitem{homes93} C.C. Homes, T. Timusk, R. Liang, D.A. Bonn, and W.N.
Hardy, Phys. Rev. Lett. {\bf 71}, 1645 (1993). \comment{ 123 pseudogap paper}

\bibitem{walstedt90} R.E. Walstedt, R.F. Bell, R.J. Cava, G.P. Espinosa,
L.F. Schneemeyer, and J.V. Waszczak, Phys. Rev B {\bf 41}, 9574 (1990).
\comment{first knight shift of pseudogap}

\bibitem{bucher93} B.~Bucher, P.~Steiner, J.~Karpinski, E.~Kaldis,
and P.~Wachter, Phys. Rev. Lett. {\bf 70}, 2012 (1993). \comment{ 124 spin gap,
normal state transport,}

\bibitem{loram94} J.W.~Loram, K.A. Mirza, J.R.~Cooper, and W.Y.~Liang,
J.of Superconductivity, {\bf 7}, 261 (1994). \comment{ \YBa specific heat
pseudogap}

\bibitem{loeser96} A.G.~Loeser, Z.-X.~Shen, D.S.~Dessau, D.S.~Marshall,
C.H.~Park, P.~Fornier, and A.~Kapitulnik, Phys. Rev. Lett. {\bf 76}, 4841
(1996).

\bibitem{ding96} H.~Ding, T.~Tokoya, J.C.~Campuzono, T.~Takahashi,
M.~Randeria, M.R.~Norman, T.~Mochiku, K.~Kadowaki, and J.~Giapitzakis, Nature
{\bf 382}, 51 (1996).

\bibitem{marshall96} D.S.~Marshall, D.S.~Dessau, A.G.~Loeser,
C.H.~Park,Z.-X.~Shen, A.Y.~Matsuura, J.N.~Eckstein, I.~Bozovik,
P.~Fornier,A.~Kapitulnik, W.E.~Spicer, and Z.-X.~Shen, Phys. Rev. Lett. {\bf
76}, 4841 (1996).
 \comment{ photoemssion \BiSr d-wave pseudogap}

\bibitem{tao97} H.J.~Tao, F.~Lu, and E.L.~Wolf, Physica C {\bf 282-287} 1507
(1997). \comment{first tunnelling pseudogap}

\bibitem{renner98} Ch.~Renner, B.~Revaz, J-Y~Genoud,
K.~Kadowaki, and O.~Fischer Phys. Rev. Lett. {\bf 80}, 149, (1998).

\bibitem{krasnov00} V.M.~Krasnov, A.~Yurgens, D.~Winkler,
P.~Delsing, and T.~Claeson, Phys. Rev. Lett. {\bf 84}, 5860 (2000).

\bibitem{slakey90} F.~Slakey, M.V.~Klein, J.P.~Rice,and
D.M.~Ginsberg, Phys. Rev B {\bf 42}, 2643 (1990). \comment{ \YBa Raman,
electronic scattering, structure present in the normal  state}

\bibitem{hackl96} R.~Hackl, G.~Krug, R.~Nemetshcek, M.~Opel, and B.~Stadlober, in
{\it Spectroscopic Studies of Supercondcutors V}, Ivan Bozovic  and Dirk van
der Marel, Editors, Proc. SPIE {\bf 2696}, 194 (1996).\comment{Superconducting
gap and pseudogap in Raman scattering.}

\bibitem{naeini99} J. G. Naeini, X. K. Chen, J. C. Irwin, M. Okuya, T. Kimura, and
K. Kishio Phys. Rev. B {\bf 59},  9642 (1999). \comment{Doping dependence of
the pseudogap in La2 - xSrxCuO4, Raman \LaSr}

\bibitem{rossat-mignod91} J.~Rossat-Mignod {\it et al.}\  Physica C
{\bf 185-189}, 86 (1991). \comment{ spin gap by neutron scattering  P.~Bourges,
P.~Burlet,}

\bibitem{tranquada92} J.M. Tranquada {\it et al.}\  Phys. Rev B {\bf 46}, 5561
(1992). \comment{ \YBa 6.6 neutron scattering, Tc=35 K, 5-50 meV 10 - 100 K 27
meV peak in Chi at 10 K Chi has a temperature independent gap at 9meV,
substantial scattering below the  BCS weak coupling gap energy (16 meV)}





\bibitem{timusk95} T.~Timusk, C.C.~Homes, and
W.~Reichardt,  Int. Workshop on the Anharmonic Properties of High T$_c$
Cuprates, Bled, Slovenia, G. Ruani, editor (World Scientific, Singapore, 1995).
\comment{The role of c-axis phonons in high temperature superconductors.}


\bibitem{homes95} C.C.~Homes, T.~Timusk, R.~Liang, D.A.~Bonn,
andW.H.~Hardy,  Can. J. Phys. {\bf 73}, 663, (1995). \comment{ 29 c-axis phonon
paper "Optical phonons along the c axis of YBa$_{2}$Cu$_{3}$O$_{6+x}$, for
${\bf x}= {\bf 0.6}\rightarrow {\bf 0.95}$}


\bibitem{homes95a} C.C.~Homes, T.Timusk, R.~Liang, D.A.~Bonn,
and W.H.~Hardy, Physica C, {\bf 254,} 265-280, (1995). \comment{ 24 c-axis
electronic properties, "Optical properties along the c axis of
YBa$_{2}$Cu$_{3}$O$_{6+x}$, for ${x}= {0.6}\rightarrow {0.95}$ "}

\bibitem{schutzmann94} J.~Sch\"utzmann, S.~Tajima, S.~Miyamoto, and
S.~Tanaka,Phys. Rev. Lett. {\bf 73}, 174 (1994). \comment{ \YBa c-axis single
crystal sigma dc=200 tc=89 K, overdoped }

\bibitem{bernhard98} C.~Berhard, R.~Henn, A.~Wittlin, M.~Kl\"aser,
Th.~Wolf, G.~M\"uller-Vogt, C.T.~Lin, and M.~Cardona, Phys. Rev.
Lett. {\bf 80}, 1762, (1998). \comment{\YBa Ca doped, overdoped c-axis,
ellipsometry, Electronic c-axis Response of Y12xCaxBa2Cu3O72d Crystals Studied
by Far-Infrared Ellipsometry}



\bibitem{kumar92} N.~Kumar and A.M.~Jayannavar, Phys. Rev. B {\bf
50}, 438 (1992).

\bibitem{chakravarty93} S. Chakravarty {\it et al.} Science {\bf 261}, 337 (1993).
\comment{theory of c-axis transport}

\bibitem{rojo93} A.G.~Rojo and K.~Levin, Phys. Rev. B {\bf 48}, 16861
(1993).

\bibitem{clarke95} D.G.~Clarke, S.P.~Strong, and P.W.~Anderson,
Phys. Rev. Lett. {\bf 74}, 4499 (1995). \comment{ inter luttinger liquid
transport}


\bibitem{abrikosov96} A.A. Abrikosov, Phys. Rev B {\bf 54}, 12003 (1996).
\comment{ pseudogap theory, c-axis conductivity resonant tunneling}

\bibitem{atkinson99} W. A. Atkinson, Phys. Rev. B {\bf  59}, 3377 (1999).
\comment{ Disorder and chain superconductivity in \YBa delta }


\bibitem{koch90b} B.~Koch, M.~D\"urrler, H.P.~Geserich,
Th.~Wolf,G.~Roth, and G.~Zachmann, in {\it Electronic Properties of
High-T$_c$Superconductors and Related Compounds,} ed. H.~Kuzmany,
M,~Mehrig,J.~Fink (Springer Series in Solid State Sciences, Vol
99,Springer-Verlag, Berlin - Heidelberg, 1990), p. 290.

\bibitem{osafune99} T. Osafune, N. Motoyama, H. Eisaki, S. Uchida,
and S. Tajima Phys. Rev. Lett {\bf 82}, 1313 (1999). \comment{Pseudogap and
Collective Mode in the Optical Conductivity Spectra of Hole-Doped Ladders in
Sr14 - xCax Cu24O41 }

\bibitem{bernhard00a} C. Bernhard, T. Holden, A. Golnik, C. T. Lin, and M.
~Cardona, Phys. Rev. B {\bf 61}, 618 (2000). \comment{c-axis far infrared
conductivity of Pr doped YBCO, also Zn doped}


\bibitem{burns91} G. Burns, F.H. Dacol, C.~Feild, F.~Holtzberg
Physica C {\bf 181}, 37 (1991). \comment{ \YBa Raman oxygen content 600 mode
from sticks}


\bibitem{vandermarel96} D. van der Marel and A.A.~Tsvetkov, Czech. J. Phys.  {\bf 46},
3165, (1996). \comment{Interlayer transverse plasmon, Theory}

\bibitem{munzar99} D. Munzar, C.~Bernhard, A.~Golnik, J.~Huml\'{\i}\v{c}ek, and
M.~Cardona, Solid State Comm. {\bf 112}, 365 (1999).

\bibitem{bernhard00b} C. Bernhard, D. Munzar, A. Golnik, C. T. Lin,
A. Wittlin, J. Huml\'{\i}\v{c}ek, and M. Cardona Phys. Rev. B {\bf
62}, 9138 (2000). \comment{Anomaly of oxygen bond-bending mode at
320  cm–1 and additional absorption peak in the c-axis infrared
conductivity of underdoped
     YBa2Cu3O7– single crystals revisited with ellipsometric measurements}

\bibitem{timusk03} T. Timusk and C.C. Homes, Solid
State Communications, (in press)

\bibitem{bernhard99} C. Bernhard, D. Munzar, A. Wittlin, W.
K\"onig, A. Golnik, C. T. Lin, M. Kl\"aser, Th. Wolf, G. M\"uller-Vogt, and M.
Cardona  Phys. Rev. B {\bf 59},  R6631 (1999).\comment{Far-infrared
ellipsometric study of the spectral gap in the c-axis conductivity of
Y12xCaxBa2Cu3O72 d crystals, Ca doped, overdoped}


\bibitem{andersen94} O.K.~Andersen, O.~Jepsen, A.J.~Licchtenstein, and
I.I.~Mazin, Phys. Rev. B. {\bf 49}, 4145 (1994). \comment{band structure
calculation showing c-axis transport at antinodal points, cited in Ioffe and
Millis prb 98}

\bibitem{andersen94b} O.K.~Andersen, J. Phys. Chem. Solids {\bf 56}, 1573
(1995). \comment{c-axis transfer integrals, quoted in Bernhard99}

\bibitem{xiang96} T.~Xiang and J.M.~Wheatley, Phys. Rev. Lett. {\bf 77}, 4632
(1996). \comment{c-axis transport theory, quoted in Berhard99}



\bibitem{puchkov96c} A.V.~Puchkov, P.~Fournier, D.N.~Basov,
T.~Timusk, A.~Kapitulnik, N.N.~Kolesnikov, Phys. Rev. Lett. {\bf 77}, 3212,
(1996). \comment{#44 BiTl2 Evolution of the Pseudogap of High-{\it T$_c$}
Superconductors with doping}



\bibitem{allen77} J.W. Allen and J.C. Mikkelsen, Phys. Rev B {\bf 15},
2952 (1977).

\bibitem{mori65} H. Mori, Prog. Theor. Phys. {\bf 34}, 399 (1965).

\bibitem{gotze72} W.~G\"otze and P.~W\"olfe, Phys. Rev. B {\bf 6}, 1226 (1972).

\bibitem{shulga91} S.V.~Shulga, O.V.~Dolgov, and E.G.~Maksimov, Physica
C \textbf{178}, 266 (1991).

\bibitem{singley01} E.J.~Singley, D.N.~Basov, K.~Kurahashi, T.~Uefuji, and
K.~Yamada, Phys. Rev. B {\bf 64}, 224503 (2001). \comment{NdCe ab plane and c
axisDiagram of oscillator fits}

\bibitem{tu01} J.J. Tu, C.C. Homes, G.D. Gu, D.N. Basov, S.M. Loureiro, R.J. Cava, M.
Strongin, Phys. Rev. B {\bf 66}, 144514 (2002).
%

\bibitem{tanner92} D.B.~Tanner and T.~Timusk, Optical
Properties of High-Temperature Superconductors, in {\it Physical Properties of
High Temperature Superconductors III} D.M.~Ginsberg, editor, (World Scientific,
Singapore, 1992) 105 pp.

\bibitem{gao93}F.~Gao, D.B.~Romero,
D.B.~Tanner, J.~Talvacchio, and M.G.~Forrester, Phys. Rev. B {\bf 47}, 1036
(1993).

\bibitem{quijada99} M.A. Quijada, D.B. Tanner, R.J.~Kelley, M.~Onellion,
H.~Berger, and G.~Maragaritondo, Phys. Rev. B {\bf 60}, 14917 (1999).
\comment{Bi-2212, anisotropy, single domain, }

\bibitem{puchkov96d} A.V.~Puchkov, D.N. Basov, and T. Timusk,   J. Physics,
Condensed Matter, {\bf 8}, 10049, (1996).
\comment{
Infrared Study,}


\bibitem{startseva99u} T.~Startseva, R.A.~Hughes, A.V.~Puchkov, D.N.~Basov,
T.~Timusk, and H.A.~Mook,  Phys. Rev. B {\bf 59}, 7184, (1999). \comment{The
temperature evolution of the pseudogap state in the infrared response of
underdoped La$_{2-x}$Sr$_x$CuO$_4$.}

\bibitem{startseva99o} T~Startseva, T.~Timusk, M.~Okuya, T.~Kimura, and
K.~Kishio,
 Physica C. {\bf 321}, 135, (1999). \comment{ LsSr, IR single crystal,
overdoped, pseudogap}

\bibitem{dumm02} M.~Dumm, D.N.~Baosv, S.~Komiya,~Y.~Abe, and Y.~Ando, Phys.
Rev. Lett. {\bf 88}, 147003 (2002).\comment{LaSr, stripes, ab planeN,dCe}


\bibitem{mcguire00}J.J. McGuire, M. Windt, T. Startseva, T. Timusk, D.
Colson, V. Viallet-Guillen,  Phys. Rev. B \textbf{62}, 8711
(2000).\comment{Pseudogap in the IR Response of
HgBa$_2$Ca$_2$Cu$_3$O$_{8+\delta}$,}



\bibitem{sacuto98} A.~Sacuto, R.~Combescot, N.~Bontemps and C.A.~M\"uller
Phys. Rev. B {\bf 58} 11721 (1998).\comment{Raman in Hg-1223}

\bibitem{julien96}M.-H.~Julien, P.~Carretta, M.~Horvati\'c,
C.~Berthier, Y.~Berthier, P.~S\'egransan, A.~Carrington, and D.~Colson Phys.
Rev. Lett. {\bf 76}, 4238, (1996). \comment{ Spin gap in HgBa2Ca2Cu3O[sub [bold
8+ delta ]] single crystals from 63Cu NMR}

\bibitem{millis93} A.J.~Millis and H.~Monien, Phys. Rev. Lett. {\bf
70}, 2810 (1993).

\bibitem{ohsugi91} S.~Ohsugi,Y.~Kitaika, K.~Ishida, and K.~Asayama, J. Phys.
Soc. Japan, {\bf 60}, 2351 (1991). comment{weak pseudogap in HTSC La214}

\bibitem{batlogg94} B.~Batlogg, H.Y.~Hwang, H.~Takagi, R.J.~Cava, H.L.
Kao, and J.~Kwo, Physica C {\bf 235--240}, 130 (1994).

\bibitem{basov95c} D.N.~Basov, H.A.~Mook, B.~Dabrowski, and T. Timusk,
Phys. Rev. B {\bf 52}, R13141 (1995). \comment{c-axis \YBa LaSr compared, low
frequency mode. Zn doping}

\bibitem{uchida96} S.~Uchida, K.~Tamasaku, and S.~Tajima,
Phys. Rev. B {\bf 53}, 14558 (1996).

\bibitem{reedyk92} M.~Reedyk and T.~Timusk, Evidence for ab-plane
Coupling to Longitudinal c-axis Phonons in High-$T_c$ CuprateSuperconductors,
Phys. Rev. Lett. 69, 2705, 1992

\bibitem{tajima} S.~Tajima, (private communication)

\bibitem{varma89} C.M.~Varma, P.B.~Littlewood, S.~Schmitt-Rink,
E.~Abrahams, and A.E.~Ruckenstein, Phys. Rev. Lett. {\bf 63}, 1996 (1989).

\bibitem{gurvitch87} M. Gurvitch and A.T. Fiory, Phys. Rev. Lett.
{\bf 59}, 1337, 1987 \comment{ dc resitivity}


\bibitem{schlesinger90} Z.~Schlesinger, R.T.~Collins, F.~Holtzberg,
C.~Feild, S.H.~Blanton, U. Welp, G.W.~Crabtree, Y.~Fang, and J.Z.~Liu, Phys.
Rev. Lett. {\bf 65}, 801 (1990). \comment{``Superconducting Energy Gap and
Normal-StateConductivity of a Single-Domain, MFL fit of scattering rate}


\bibitem{Ts temperature} It is clear that the gap in the scattering rate
develops below 200 K and above the onset of superconductvity $T_c$ but there is
an unfortunate lack of data between these two temperatures to determine the
exact onset temperature of MFL scattering.

\bibitem{schachinger01} E.~Schachinger and J.P.~Carbotte, Phys. Rev. B {\bf
64}, 094501 (2001). \comment{spin fluctuation spectrum data of Homes}

\bibitem{bourges98} For a review of the resonance peak in neutron scattering see
P.~Bourges, in \textit{The gap Symmetry and Fluctuations in High Temperature
Superconductors}, ed. J.~Bok, G.~Deutscher, D.~Parrina, and S.A.~Wolf, Plenum
Press 1998, preprint cond-mat/9901333.

\bibitem{tanner private} D.A.~Tanner, private communication.

\bibitem{eschrig00} M.~Eschrig and M.R.~Norman, Phys. Rev. Lett. {\bf 85}, 3261
(2000).\comment{caluction of ARPES dispersion}

\bibitem{kaminski00} A.~Kaminski, J.~Mesot, H.~Fretwell, J.C.~Campuzano,
M.R.~Norman, M.~Randeria, H.~Ding, T.~Sato, T.~Takahashi, T.~Mochiku,
K.~Kadowaki, and H.~Hoechst, Phys. Rev. Lett. \textbf{84}, 1788 (2000).
\comment{Evidence for an energy scale for quasiparticle dispersion in
Bi2Sr2CaCu2O8. (preprint cond-mat/0004349)}

\bibitem{bogdanov00} P.V.~Bogdanov, A.~Lanzara, S.A.~Kellar, X.J.~Zhou, E.D.~Lu,
W.J.~Zheng, G.~Gu, J.-I.~Shimoyama, K.~Kishio, H.~Ikeda, R.~Yoshizaki,
Z.~Hussain, and Z.X.~Shen, Phys. Rev. Lett. \textbf{85}, 2581 (2000).
\comment{Evidence for an Energy Scale for Quasiparticle Dispersion in
Bi2Sr2CaCu2O8}

\bibitem{norman99} M.R.~Norman, H.~Ding, H.~Fretwell, M.~Randeria, J.C.~Campuzano,
Phys. Rev. B \textbf{60}, 7585 (1999). \comment{Evidence for an energy scale
for quasiparticle dispersion in Bi2Sr2CaCu2O8. Comparsion with optics}

\bibitem{lanzara01} A. Lanzara et al. Nature \textbf{412}, 510
(2001).\comment{full author list: A. Lanzara, P.V.~Bogdanov, X.J.~Zhou,
S.A.~Keller, D.L.~Feng, E.D.~Lu, Y.~Yoshida, A.~Fujimori, K.~Kishio,
J.-I.~Shimoyama, T.~Noda, S.~Uchida, Z.~Hussain, and Z._X.~Shen}

\bibitem{wang02} N.L.~Wang, T. Timusk, J.P. Franck, P. Schweiss, M. Braden, A. Erb,
The oxygen isotope effect in the ab-plane reflectance of underdoped
YBa$_2$Cu$_3$O$_{7-\delta}$, Phys. Rev. Letters {\bf 89}, 087003 (2002).


\bibitem{fong99} H.F.~Fong, P.~Bourges, Y.~Sidis, L.P.~Regnault, J.~Bossy,
 A.~Ivanov, D.L.~Milius, I. A.~Aksay, and B.~Keimer, Phys. Rev. Letters {\bf 82}, 1939 (1999).
 \comment{Effect of Nonmagnetic Impurities on the Magnetic Resonance Peak in YBa2Cu3O7, zinc}

\bibitem{zheng93} G.-q~Zheng, T.~Odaguchi, T.~Mito, Y.~Kitaoka, K.~Asayama,
and Y.~Kodama, J. Phys. Soc. Japan {\bf 62}, 2591 (1993).

\bibitem{zheng96} G.-q~Zheng, T.~Odaguchi, Y.~Kitaoka, K.~Asayama, Y.~Kodama,
K.~Mizuhashi, and S.~Uchida, Physica C {\bf 263} 367 (1996).

\bibitem{nmr gaps} For a review of the experimental NMR data on the pseudogap
in high temperature superconductors see T. Timusk and B. Statt, Reports of
Progress in Physics {\bf 62}, 61 (1999).

\bibitem{basov94c} D.N.~Basov and T.~Timusk, B.~Dabrowski and
J.D.~Jorgensen, Phys. Rev B {\bf 50}, 3511 (1994). \comment{C-axis Response of
YBa$_2$Cu$_4$O$_8$ Single Crystals in Normal and Superconducting State.}


\bibitem{harris96} J.M.~Harris, A.G.~Loeser, D.S.~Marshall, M.C.~Schabel, and Z.-X.~Shen Phys.
Rev. B {\bf 54} R15665 (1996).


\bibitem{erb_liang} A.~Erb, E.~Walker, and R.~Flukiger, Physica C
{\bf 258}, 9 (1996), R.X.~Liang, D.A.~Bonn, amd W.N.~Hardy, Physica C {\bf
304}, 105 (1998).

\end{thebibliography}
\end{document}